\documentstyle[epsf,aps,preprint]{revtex}
\tightenlines

\newcommand{\lsim}{\mathrel{\lower4pt\hbox{$\sim$}}
\hskip-12.5pt\raise1.6pt\hbox{$<$}\;}
\newcommand{\gsim}{\mathrel{\lower4pt\hbox{$\sim$}}
\hskip-12.5pt\raise1.6pt\hbox{$>$}\;}

\def\be{\begin{equation}}
\def\ee{\end{equation}}
\def\bea{\begin{eqnarray}}
\def\eea{\end{eqnarray}}
\def\ba{\begin{array}{ccc}}
\def\ea{\end{array}}
\def\M.5{M^{-\frac{1}{2}}}

\begin{document}
%\baselinestretch 3
\baselineskip14pt
%\addtolength{\baselineskip}{-.3\baselineskip}

\vspace{-.5in}
\hskip4.5in
\vbox{\hbox{PURD--TH--99--09}
\hbox{OITS-685}
\hbox{hep-ph/9912366}}
%\hbox{December, 1999}}

\vspace{.5in}
\begin{center}
{\large\bf Mass Hierarchies  and the Seesaw Neutrino Mixing}

\vspace{.5in}

T. K. Kuo$^a$\footnote{Email: tkkuo@physics.purdue.edu,
$^\dag$ wu@dirac.uoregon.edu, $^\ddag$ mansour@physics.purdue.edu},
Guo-Hong Wu$^{b\dag}$, and Sadek W. Mansour$^{a\ddag}$

\vspace{.3in}
{\it $^a$Department of Physics, Purdue University,
        West Lafayette, IN 47907\\}
\medskip
{\it $^b$ Institute  of Theoretical Science,
University of Oregon, Eugene, OR 97403}

\vspace{.3in}
{Revised, Jan 2000}

\end{center}
%\bigskip

%\baselineskip24pt
%\baselineskip16pt

\begin{abstract} 

We give a general analysis of neutrino mixing in the seesaw mechanism 
with three
flavors. Assuming that the Dirac and u-quark mass matrices are similar, we
establish simple relations between the neutrino parameters and
individual Majorana masses. They are shown to depend rather strongly on the
physical neutrino mixing angles. We calculate explicitly the implied Majorana 
mass hierarchies for parameter sets corresponding to different solutions to 
the solar neutrino problem.

\end{abstract}

\newpage
\tightenlines

\section{Introduction}

One of the most pressing questions in particle physics has been the
 determination of the intrinsic properties of the neutrinos, namely, 
their masses and mixing angles.  The recent atmospheric neutrino 
data~\cite{atmos} suggest strongly that neutrinos do have 
masses, and that, unlike the quark sector, at least some of the mixing 
angles are large.
The most appealing model for small neutrino masses derives from the seesaw 
mechanism~\cite{ss}.    
In doing so, however, one also introduces additional unknowns in the form 
of the Dirac and Majorana mass matrices. We do have a handle on the Dirac 
matrix, since, from the ideas of GUTs, it should be similar to that of the 
quark sector.  Not much is known about the Majorana mass matrix. The 
challenge is then to find out what the Majorana matrix is like in order 
for the effective neutrino mass matrix to come out correctly.

  In this paper, we will analyze the general structure of the seesaw 
mass matrix.  Without loss of generality, we will work in the basis 
in which the charged lepton and Majorana  mass matrices are
diagonal.
As a starting point, we assume that the Dirac mass matrix, in analogy to the
quark mass matrix, has hierarchical eigenvalues and small left-handed
mixing angles. 
Even in this case, large mixing can occur through the interplay of the
Dirac and Majorana matrices~\cite{ssen,tanimoto,afm,branco,jez}.

 In the seesaw mechanism, the effective neutrino mass matrix is given by
\begin{eqnarray}
m^{\rm eff} & = & m_D M^{-1} m_D^T
\end{eqnarray}
The Dirac matrix can be written as 
\begin{eqnarray}
m_D & = & U_0 m_D^{\rm diag} V_0
\end{eqnarray}
in the basis where $M^{-1}$ is diagonal,
\begin{eqnarray}
M^{-1} & = & \left( \matrix{
                 R_1^2 & 0 & 0 \cr
                 0 & R_2^2 & 0 \cr
                 0 & 0 & R_3^2 } \right)
\end{eqnarray}
Here $U_0$ and $V_0$ are left-handed (LH) and right-handed (RH) rotations, 
respectively.
$m_D^{\rm diag}$ is a diagonal matrix with eigenvalues
$m_i$ ($i=1,2,3$). The Majorana mass eigenvalues are
given by $M_i=1/R^2_i$ ($i=1,2,3$).
For simplicity, we assume that $U_0$ and $V_0$ are real (i.e. we ignore 
CP violating
effects). Note that $V_0$ also contains contribution from the diagonalization
of $M^{-1}$, and thus it may contain large angles.  
However, we will restrict our discussions to small angles in $V_0$.

  It is convenient to write 
\begin{eqnarray}
U^{-1}_0 m^{\rm eff} U_0 & = & N N^T
\end{eqnarray}
where 
\begin{eqnarray} \label{eq:nmatrix}
N& =& m_D^{\rm diag} V_0 M^{-\frac{1}{2}} 
   =   \left( \matrix{
           m_1 & & \cr
              & m_2 & \cr
            & & m_3 } \right)
         V_0 \left( \matrix{
              R_1 & & \cr
              & R_2 & \cr
            & & R_3 } \right) \; .
\end{eqnarray}
For hierarchical Dirac masses, to a good approximation
\begin{eqnarray}
\left( \matrix{
           m_1 & & \cr
              & m_2 & \cr
            & & m_3 } \right) V_0 
 & \simeq & \left( \matrix{
            m_1 V_{11} & 0 & 0 \cr
            m_2 V_{21} & m_2 V_{22} & 0 \cr
            m_3 V_{31} & m_3 V_{32} & m_3 V_{33} } \right) \; .
\end{eqnarray}
Here $V_{ij} \equiv (V_0)_{ij}$ are the matrix elements of $V_0$.
More precisely~\cite{triang}, the leading correction to this approximation is a 
LH rotation with rotation angles 
$(\phi_{12}, \phi_{13}, \phi_{23}) \simeq 
 \left( \frac{m_1}{m_2}\frac{V_{12}}{V_{22}},
        \frac{m_1}{m_3}\frac{V_{13}}{V_{33}},
        \frac{m_2}{m_3}\frac{V_{23}}{V_{33}} \right)$ .
This can be absorbed into $U_0$, and will be ignored henceforth.

      If $M^{-1} \propto I$, then $NN^T$ will be diagonal and
$m^{\rm eff}$ will be diagonalized by $U_0$.  
 When $M^{-1}$ deviates from the identity matrix, the product 
$V_0M^{-\frac{1}{2}}$ is no longer orthogonal, then
\be\label{eq:Ndiag}
N = U N^{\rm diag} W 
\ee
where $U$ and $W$ are LH and RH rotations respectively.
The LH rotation $U$ is induced by $M^{-\frac{1}{2}}$ so that
\be\label{eq:U0}
(U^{-1}U_0^{-1})\ m^{\rm eff} \ (U_0U)= (N^{\rm diag})^2
\ee
Thus, the mixing angles for the effective neutrino mass matrix come from 
$U_0U$, while the eigenvalues of $N$ are just $\sqrt{m^{\rm eff}_i}$, with
$m^{\rm eff}_i$ denoting the effective neutrino masses.
To the extent that we may assume $U_0$ to be nearly the identity matrix,
the case of large mixing angles receives its main contribution from $U$, 
which is induced by $M$.

  From Eq.~(\ref{eq:nmatrix}), we have
\bea \label{eq:TD}
N & \simeq & \left( \matrix{ 
       R_1m_1V_{11} & 0 & 0 \cr
       R_1m_2V_{21} & R_2m_2V_{22} & 0 \cr
       R_1m_3V_{31} & R_2m_3 V_{32} & R_3 m_3 V_{33} } \right)\; .
\eea
While we have assumed that the Dirac mass eigenvalues $m_i$ have the
hierarchical structure of the quark masses, little is known about the
values of $R_i$. 

The problem within the seesaw model is then twofold. One, given the 
parameters $m_i$, $R_i$, and $V_{ij}$, what is the LH rotation matrix
$U$ as defined in Eq.~(\ref{eq:Ndiag}) ? Conversely, if we know $U$, i.e.
the physical mixing angles for the neutrinos, what can we deduce about
these parameters? We will discuss these issues in the following sections.

\section{General Analysis}
Let us start from a general
$3\times 3$ matrix
\be\label{eq:Ngeneral}
N=\left(\matrix{a_1&a_2&a_3\cr b_1&b_2&b_3\cr c_1&c_2&c_3}\right)\; .
\ee
Anticipating the applications to the neutrino sector, we will assume
\be
{a}^2\ll {b}^2\sim {c}^2\; ,
\ee
 where we have used the notation $\vec{a}=(a_1,a_2,a_3)$, etc. 
We have shown elsewhere~\cite{triang} that when a matrix is brought into the
upper triangular form, we can obtain the LH mixing angles easily. Now, it is 
always possible to find a RH rotation so that $N$ becomes
upper triangular. In fact, geometrically, this amounts to a new coordinate
system where $\vec{c}$ is aligned with the third axis, while the second
and first axes are in the directions $\vec{c}\times(\vec{b}\times \vec{c})$
and $\vec{b}\times \vec{c}$, respectively. It is then clear that, for an
appropriate RH rotation $R$, 
\be\label{eq:Nt}
N_t= N R = \left(\matrix{\vec{a}.\hat{i}& \vec{a}.\hat{j}& \vec{a}.\hat{c}\cr
0&|\vec{b}\times\hat{c}|&\vec{b}.\hat{c}\cr 0&0&c}\right)\; ,
\ee
where $\hat{i}=\vec{b}\times\vec{c}/|\vec{b}\times\vec{c}|$,
$\hat{j}=\vec{c}\times(\vec{b}\times\vec{c})/(|\vec{c}||\vec{b}\times\vec{c}|)$,
and $\hat{c}=\vec{c}/|\vec{c}|$.
Note that, since the physical mass matrix is given by $N N^T$, the matrix $N$
in Eq.~(\ref{eq:Ngeneral}) is arbitrary up to any RH rotation.
The matrix in Eq.~(\ref{eq:Nt}), however, is constructed entirely from 
rotational
invariants, and is given in terms of physical quantities. There is still some
ambiguity in Eq.~(\ref{eq:Nt}), in that the diagonal elements can be of 
either sign,
corresponding to the choice of orientation of the axes.

To diagonalize $N$, we first diagonalize its (23) submatrix explicitly by
$R_L(23)$ and $R_R(23)$, with
\bea\label{eq:23LR}
\tan 2\theta_{23}^L&=&\frac{2 |\vec{b}.\vec{c}|}{c^2-b^2} \nonumber\\
\tan 2\theta_{23}^R&=&\frac{2 |\vec{b}.\vec{c}| |\vec{b}\times\vec{c}|}
{c^4+(\vec{b}.\vec{c})^2-(\vec{b}\times \vec{c})^2} \; .
\eea 
Further, the eigenvalues of $N$ are $\mu_2$ and $\mu_3$, with
\bea\label{eq:eigenvalues}
\mu_{2,3}^2&=&\frac{1}{2}(c^2+b^2)\mp \Delta \nonumber\\
\Delta^2&=&\frac{1}{4}(c^2+b^2)^2-|\vec{b}\times\vec{c}|^2\; .
\eea
Thus we find
\be
N_t'=R_L(23) N_t R_R(23)=\left(\matrix{\vec{a}.\hat{i}& \alpha_2& \alpha_3 \cr
0&\mu_2&0\cr 0&0&\mu_3}\right)\; ,
\ee
where $(\alpha_2, \alpha_3)=(\vec{a}.\hat{j}, \vec{a}.\hat{c})R_R(23)$.
Note that the mixing angle $\theta_{23}^L$ is maximal when $b^2 = c^2$.
Also, a mass hierarchy, $\mu_3\gg \mu_2$, implies that 
$\vec{b}.\vec{c}\approx c^2$,
i.e. $\vec{b}$ and $\vec{c}$ are nearly parallel.
Since $|\alpha_3|\lsim a\ll \mu_3$, a LH (13) rotation $R_L(13)$, 
with $\theta_L^{13}\sim
|\alpha_3|/\mu_3\ll 1$, removes the (13) element of $N_t'$ without changing
the other elements appreciably.
Summarizing, we see that, if $a^2\ll c^2$,
\be
\bar{N}=R_L(13)R_L(23) N R R_R(23)\simeq \left(\matrix{\vec{a}.\hat{i}& 
\alpha_2& 0\cr 0&\mu_2&0\cr 0&0&\mu_3}\right)\; ,
\ee
where $R_L(13)\simeq I$, $R$ is defined in Eq.~(\ref{eq:Nt}), and the rotation 
angles in $R_{L,R}(23)$ are given in Eq.~(\ref{eq:23LR}).
The final diagonalization of $\bar{N}$ can be achieved by a combined LH and RH
(12) rotation. In particular, maximal mixing is possible if 
$\alpha_2\sim \mu_2$.
We emphasize that as long as the (13) rotation $R_L(13)$ is small, the 
diagonalization of $N$ can be decomposed into that of two $2\times 2$ 
matrices. This is a very useful simplification.

Of particular interest in neutrino physics is the possibility of large 
mixing angles. Assuming
$|\vec{a}|\ll|\vec{c}|$, which ensures that $\theta_{13}^L\ll 1$, 
maximal (23) rotation, 
$\theta_{23}^L = \pi/4$, is obtained if $|\vec{b}| = |\vec{c}|$. 
There are now two possibilities:\\
A) $|\vec{b}.\vec{c}|\gg |\vec{b}\times \vec{c}|^2.$\\
In this case, from Eq.~(\ref{eq:eigenvalues}) we find that
$
\mu_2^2\simeq |\vec{b}\times\vec{c}|^2/(c^2+b^2)\ll \mu_3^2\; .
$
Under this condition, we can have bimaximal mixing, $\theta^L_{12}\simeq \pi/4$. 
This is achieved if $\alpha_2\simeq \mu_2$. Note that, for 
$|\vec{b}\times\vec{c}|\ll |\vec{b}.\vec{c}|$, we have the bound 
$|\tan 2 \theta_R^{23}|\lsim |\vec{b}\times\vec{c}|/|\vec{b}.\vec{c}|$.
Thus $|\sin \theta_R^{23} \vec{a}.\hat{c}|\ll \mu_2$,
so the condition $\alpha_2\simeq \mu_2$ can also be written 
as $\vec{a}.\hat{j}\simeq \mu_2$.\\
B) $|\vec{b}.\vec{c}|\sim |\vec{b}\times\vec{c}|.$\\
In this case $\mu_2\sim \mu_3$.
All of the components of $\vec{a}$ are smaller than $\mu_2$, in particular,
$|\mu_2|\gg \alpha_2$.
Thus, $\theta^L_{12}\sim |\alpha_2|/|\mu_2|\ll 1$, and we can only have single
maximal mixing.

Knowing $N$, the above analysis yields the neutrino mixing angles 
$\theta_{ij}^L$. 
In Eq.~(\ref{eq:TD}), $N$ is given in terms of the parameters $m_i$, $R_i$, 
and $V_{ij}$.
There are generally two classes of possibilities: (A) All $R_i$ are of the 
same order, or (B) there is a strong hierarchy in the Majorana sector.

For the case of no Majorana hierarchy, $R_1\sim R_2\sim R_3$, we can find 
approximately the rotation angles, which give
\be
{\bar R}_L=R_{12}\left(\frac{m_1 R_1^2}{m_2 R_2^2}V_{21}\right)
R_{13}\left(\frac{m_1 R_1^2}{m_3 R_3^2}V_{31}\right) 
R_{23}\left(\frac{m_2 R_2^2}{m_3 R_3^2}V_{32}\right)\; ,
\ee
so that
\be
\bar{R}_L N N^T \bar{R}_L^T = N^2_{\rm diag}\; .
\ee
In other words, the induced neutrino mixing angles are given approximately by

\be
\theta_{IJ}^L\simeq \frac{m_I M_J}{m_J M_I} V_{JI}\; ,\;\;\;
(I,J)=(1,2),~(2,3),~(1,3).
\ee
Unless there is a stronger hierarchy in the Majorana sector in comparison 
to the Dirac sector, all neutrino mixing angles are negligible when 
$V_{ij}$ is reasonably small.

To analyze the situation when there is a strong hierarchy in $R_i$, 
we will limit our discussion to situations which are physically 
interesting. For this purpose, it is actually more convenient to reverse 
our procedure above, i.e., given the physical
mixing angles, we deduce the form of the matrix $N$. We note that, 
experimentally, it is quite suggestive that the neutrino masses are 
hierarchical, and that the neutrino mixing matrix $U_{\alpha i}$, 
($\alpha=e,\mu,\tau; i=1,2,3$), is given approximately by:
\bea
U&=&R_{23}(\phi)R_{13}(\epsilon)R_{12}(\theta)\nonumber\\
&\simeq&\left(\matrix{c_\theta & s_\theta & \epsilon\cr
-(s_\theta c_\phi+ \epsilon c_\theta s_\phi) & (c_\theta c_\phi- 
\epsilon s_\theta s_\phi)
& s_\phi \cr 
(s_\theta s_\phi - \epsilon c_\theta c_\phi) & -(c_\theta s_\phi + 
\epsilon s_\theta c_\phi) & 
c_\phi}\right)\; .
\eea
where $\tan\phi\sim{\cal O}(1)$ and $\epsilon\ll 1$~\cite{chooz}.  
We have kept only the leading terms of $\epsilon$ in $U$.
Here, $\tan \theta$ can be either of order 1 (corresponding to the 
``bi-maximal'' mixing scenario), or could be smaller (``single-maximal'' 
scenario).  In the latter situation, we will discuss two possibilities: 
(A) $\epsilon\ll s_\theta$, (B) $\epsilon\gg s_\theta$. Appropriate 
approximations can be obtained for each of them. 

For the case $\epsilon\ll s_\theta$, the matrix $N$ is 
\be\label{eq:NU}
N=U\left(\matrix{n_1 & 0 & 0\cr 0 & n_2 & 0\cr 0 & 0 & n_3}\right)=
\left(\matrix{c_\theta n_1& s_\theta n_2 & \epsilon n_3\cr
-s_\theta c_\phi n_1& c_\theta c_\phi n_2& s_\phi n_3\cr 
s_\theta s_\phi n_1& -c_\theta s_\phi n_2& c_\phi n_3}\right)\; .
\ee
up to an arbitrary RH rotation.
>From Eq.~(\ref{eq:NU}), with the assumption $n_1\ll n_2\ll n_3$, 
it is readily seen that
\be
\frac{b^2}{s_\phi^2}\simeq \frac{c^2}{c_\phi^2} = n_3^2 = m_3^{\rm eff}\; ,
\ee
and
\be
\frac{|\vec{b}\times\vec{c}|}{|\vec{b}.\vec{c}|}\simeq
\frac{c_\theta n_2 n_3}{s_\phi c_\phi n_3^2}=\frac{c_\theta 
n_2}{s_\phi c_\phi n_3}\ll 1 \; .
\ee
These two equations imply that $\vec{b}$ and $\vec{c}$ are nearly parallel to 
leading order in $n_2/n_3$.
Also $a\ll b,c$, so that our general analysis is applicable to $N$.

We can put Eq.~(\ref{eq:NU}) in the lower triangular form through a 
RH rotation. We have 
\bea \label{eq:Nlower}
N & = &\left(\matrix{a & 0 & 0\cr
\vec{b}.\vec{a}/a & |\vec{a}\times\vec{b}|/a & 0\cr
\vec{c}.\vec{a}/a & \frac{(\vec{a}\times\vec{b}).(\vec{a}\times\vec{c})}
{a|\vec{a}\times\vec{b}|} & \frac{(\vec{a}\times\vec{b}).
\vec{c}}{|\vec{a}\times\vec{b}|}}\right)
\nonumber\\\nonumber\\
& \simeq &\left(\matrix{\sqrt{c_\theta^2 n_1^2 + s_\theta^2 
n_2^2+\epsilon^2 n_3^2} & 0 & 0\cr
(c_\phi s_\theta c_\theta n_2^2+s_\phi \epsilon n_3^2)/a &
s_\theta s_\phi n_2 n_3 /a & 0\cr
(-s_\phi s_\theta c_\theta n_2^2+c_\phi \epsilon n_3^2)/a &
s_\theta c_\phi n_2 n_3 /a & n_1/s_\theta s_\phi}\right)\; .
\eea
Note that, as in Eq.~(\ref{eq:Nt}), the entries of $N$ in 
Eq.~(\ref{eq:Nlower}) is uniquely determined by the physical neutrino 
parameters, while Eq.~(\ref{eq:NU}) is not.

A similar analysis can be done for $\epsilon \gg s_\theta$. Here, 
Eqs.~(\ref{eq:NU}) and (\ref{eq:Nlower}) are no longer valid. 
To lowest order in $s_\theta$ and $\epsilon$ and with $\phi=\pi/4$, we obtain
\be\label{eq:Nlower2}
N\simeq \left(\matrix{\epsilon n_3 & 0 & 0\cr 
\frac{n_3}{\sqrt{2}}\left(1+\frac{s_{2\theta} 
n_2^2}{2\epsilon n_3^2}\right) & 
\frac{n_2}{\sqrt{2}}\left(c_\theta-\frac{s_\theta}{\epsilon}\right) & 0\cr
\frac{n_3}{\sqrt{2}}\left(1-\frac{s_{2\theta} n_2^2}{2\epsilon n_3^2}\right) & 
-\frac{n_2}{\sqrt{2}} \left(c_\theta+\frac{s_\theta}{\epsilon}\right) & 
\frac{\sqrt{2}n_1}{\epsilon}}\right)\; .
\ee
We may now compare Eq.~(\ref{eq:TD}) to Eqs.~(\ref{eq:Nlower}) and 
(\ref{eq:Nlower2}).  Given the neutrino parameters, $n_i$, $s_\theta$, 
and $\epsilon$, this method gives immediately the Majorana masses $M_i$, 
assuming that we may identify the Dirac masses with the quark masses. 
We emphasize that this comparison is viable because the lower triangular 
form for the $N$ matrix is unique, there being no more ambiguities due to 
RH rotations. Also, the structure in these equations shows immediately that 
they are consistent only if there is a very strong hierarchy, 
$R_1\gg R_2\gg R_3$. In the next section, we will discuss in
detail the physical consequences of this result.

%%%%%%%%%%%%%%%%%%%%%%%%%%%%%%%%%%%%%%%%%%%%%%%%%%%%%%%%%%%%%%%%%%%%%%%%%%%%%%%%%%%%%%%%%%%%%%
\section{Discussions and Numerical Results}
%%%%%%%%%%%%%%%%%%%%%%%%%%%%%%%%%%%%%%%%%%%%%%%%%%%%%%%%%%%%%%%%%%%%%%%%%%%%%%%%%%%%%%%%%%%%

Assuming hierarchical Dirac neutrino masses and small RH rotation angles, 
it is seen from Eqs.~(\ref{eq:TD}) that the $N$ matrix is naturally of 
lower-triangular form. The discussion of physical constraints can be 
done most conveniently in this lower triangular basis.
In this section, we will start from Eqs.~(\ref{eq:TD}), (\ref{eq:Nlower}), and 
(\ref{eq:Nlower2}) and present the results for the neutrino parameters 
derived from different 
solutions to the solar and atmospheric neutrino observations. 
To be concrete, we will use $\phi=45^\circ$, 
and $m_i=(m_u, m_c, m_t)$ (at the seesaw scale), 
although different choices can be accommodated. \\\\
(A) $\epsilon\ll s_\theta$:\\\\
Depending on $\epsilon$, Eq.~(\ref{eq:Nlower}) gives rise to three 
different patterns, with the largest elements
located in the (a) (2,2), (3,2), (b) $(i,j)$, $i=1,2$, (c) (2,1), (3,1) 
positions, respectively. 
It can be seen that only type (a) with 
$\epsilon\ll m_2^{\rm eff}/m_3^{\rm eff}$ is natural, 
which we will concentrate on. In this limit, with $\tan\theta > n_1/n_2$,
\be\label{eq:Nesmall}
N\simeq \left(\matrix{s_\theta n_2 & 0 &0\cr \frac{1}{\sqrt{2}} c_\theta n_2 &  \frac{1}{\sqrt{2}} n_3 & 0\cr
-\frac{1}{\sqrt{2}} c_\theta n_2 & \frac{1}{\sqrt{2}} n_3 & 
\sqrt{2}\frac{n_1}{s_\theta}}\right)\; ,
\ee
where we identify $n_i^2=m_i^{\rm eff}$.
Comparison to Eq.~(\ref{eq:TD}) immediately yields (with $V_{11}\approx 
V_{22}\approx V_{33}\approx 1$)
 \bea\label{eq:meffvalues}
m_2^{\rm eff}= n_2^2 = (R_1 m_1)^2/s_\theta^2 = \frac{m_u^2}{s_\theta^2 M_1}\; ,\nonumber\\
m_3^{\rm eff}= n_3^2 = 2(R_2 m_2)^2 =\frac{2 m_c^2}{M_2}\; ,\nonumber\\
m_1^{\rm eff}= n_1^2 = s_\theta^2 (R_3 m_3)^2/2 = \frac{s_\theta^2 
m_t^2}{2 M_3}\; .
\eea
\be\label{eq:RHsmall}
-\frac{V_{31}}{V_{21}}=\frac{V_{32}}{V_{22}}
=\frac{m_2}{m_3}=\frac{m_c}{m_t}\; .
\ee
Quite surprisingly, $m_3^{\rm eff}$ scales as $m_c^2$ rather than $m_t^2$.
This is because in Eq.~(\ref{eq:Nesmall}) the (22) element is one of 
the largest.  This gives a scale
for $M_2$ much lower than one would expect,
\be
M_2\simeq \frac{2m_c^2}{\sqrt{\Delta m_{\rm atm}^2}}
\simeq 6\times 10^9 {\rm GeV}\; ,
\ee
where we have used $m_c(M_2)\simeq 0.4$ GeV, and $m_3^{\rm eff}=\sqrt{\Delta 
m_{\rm atm}^2}= \sqrt{3\times 10^{-3}}$ eV, which will also be used in the 
following.

Note also that $V_{32}$ is independent of the Majorana mass 
$M_i$ and that it scales linearly with $m_c/m_t$. This means that the 
Majorana sector decouples owing to its very large hierarchy.
Similarly, we have
\be\label{eq:RH2}
\frac{V_{21}}{V_{11}}=\frac{m_1}{\sqrt{2}\tan \theta 
m_2}=\frac{m_u}{\sqrt{2}\tan \theta m_c}\; ,
\ee
\be\label{eq:RH3}
\frac{V_{31}}{V_{11}}=-\frac{m_1}{\sqrt{2}\tan \theta 
m_3}=\frac{- m_u}{\sqrt{2}\tan \theta m_t}\; .
\ee
>From Eqs.~(\ref{eq:RHsmall}), (\ref{eq:RH2}), (\ref{eq:RH3}), 
we see that all of the RH angles are small.

The other two heavy Majorana mass values depend on the different solutions 
to the solar neutrino problem~\cite{solarnu}.  
We will use $m_3^{\rm eff}=\sqrt{\Delta m_{\rm atm}^2}$, 
$m_2^{\rm eff}=\sqrt{\Delta m_{\rm solar}^2}$, while $m_1^{\rm eff}$ is 
not known. However, we can derive a bound for $M_3$ using the 
parameter $r=m_2^{\rm eff}/m_1^{\rm eff}\gg 1$. From 
Eq.~(\ref{eq:meffvalues}), we note that
$M_1$ and $M_3$, besides the usual Dirac mass squared, depend 
sensitively on $s_\theta$ as well as on the effective neutrino masses. This 
can be seen directly in the following equations:
\be\label{eq:Mratios}
\frac{M_1}{M_2}= \frac{m_u^2 m_3^{\rm eff}}{2 m_c^2 s_\theta^2 
m_2^{\rm eff}}\; ,\;\;\;
\frac{M_2}{M_3}= \frac{4 m_c^2 m_1^{\rm eff}}{m_t^2 s_\theta^2 
m_3^{\rm eff}}\; .
\ee
We will now turn to numerical estimates with inputs coming from the three 
solutions to the solar neutrino problem, vacuum oscillations (VO), large 
angle MSW (LAM), and small angle MSW (SAM).

{\center \em A1.~VO}\\\\
The neutrino mass and mixing can be taken as \cite{solarnu},
\be
\theta \simeq 45^\circ, \;\;\;m_2^{\rm eff}=\sqrt{\Delta M_{\rm 
solar}^2}=\sqrt{7\times 10^{-11}}~{\rm eV}.
\ee
>From Eq.~(\ref{eq:Mratios}), we find
\be
M_1=\frac{2 m_u^2}{m_2^{\rm eff}}\approx 5\times 10^8~{\rm GeV}\; ,
\ee
\be
M_3/r \approx 4\times 10^{17}~{\rm GeV}\;\;\;(r\equiv 
m_2^{\rm eff}/m_1^{\rm eff}\gg 1)\; .
\ee

{\center \em A2.~LAM}\\\\
Here we take \cite{solarnu},
\be
\sin^2 2\theta_{\rm solar}\simeq 0.8\; ,\;\;\;m_2^{\rm 
eff}\approx\sqrt{\Delta m^2_{\rm solar}}=\sqrt{3\times 10^{-5}}~{\rm eV}\; .
\ee
Going through the same analysis as in the VO case, we have:
\be
M_1\approx 1\times 10^6~{\rm GeV}\; ,\;\;M_2\approx 6\times 
10^9~{\rm GeV}\; ,\;\;M_3/r\approx 4\times 10^{14}~{\rm GeV}\;\;\;(r\gg 1)\;  .
\ee

{\center \em A3.~SAM-I}\\\\
We now have  \cite{solarnu}
\be
\sin^2 2\theta_{\rm solar}\simeq 5\times 10^{-3}\; ,\;\;\;m_2^{\rm eff}\approx
\sqrt{\Delta m^2_{\rm solar}}=\sqrt{5\times 10^{-6}}~{\rm eV}\; .
\ee
They imply
\be
M_1\approx 7\times 10^8~{\rm GeV}\; ,\;\;M_2\approx 
6\times 10^9~{\rm GeV}\; ,\;\;M_3/r\approx 4\times 
10^{12}~{\rm GeV}\;\;\;(r\gg 1)\;  .
\ee

The above results were obtained under the assumption that 
$\epsilon\rightarrow 0$. Thus, if $\epsilon \gg n_2/n_3$, 
Eq.~(\ref{eq:Nesmall}) is no longer valid.
Also, if $\epsilon\gg s_\theta$, Eq.~(\ref{eq:Nlower}) should be 
replaced by Eq.~(\ref{eq:Nlower2}).
This condition is actually very likely to be valid for the case of the 
SAM solution. We treat this case in detail next.

{\center (B)~$\epsilon\gg s_\theta$ {\em (SAM-II)}:}\\\\
The parameters for $\theta$ and $m_2^{\rm eff}$ are taken as in SAM-I. 
However, we now use Eq.~(\ref{eq:Nlower2})
instead of Eq.~(\ref{eq:Nlower}). Also, the numerical result depends on 
the value of $\epsilon$, which, for 
definiteness, we will take to be $\epsilon=0.1$. We obtain
\be
M_1\approx 2\times 10^6~{\rm GeV}\; ,\;\;M_2\approx 1\times 
10^{11}~{\rm GeV}\; ,\;\;
M_3/r\approx 2\times 10^{13}~{\rm GeV}\;\;\;(r\gg 1)\;  .
\ee
\vspace{0.2 in}

In summary, we find that if the physical neutrino parameters are known, 
we can obtain the Majorana masses directly when we identify the Dirac 
masses with the quark masses. The Majorana masses have rather strong
dependence on the physical mixing angles, so that their values span 
a wide range. Numerically, it is noteworthy that 
$M_2\approx 6\times 10^9~{\rm GeV}$ for a wide range of parameters. 
Also, the VO solution for $M_3~(\gg 4\times
10^{17}~{\rm GeV})$ seems too large for it to be viable.
Finally, in some cases $M_1$ can be rather low ($\sim 10^6~{\rm GeV}$).
%%%%%%%%%%%%%%%%%%%%%%%%%%%%%%%%%%%%%%%%%%%%%%%%%%%%%%%%%%%%%%%%%%%%%%%%%%%%%%%%%%%%%%%%%%%%
\section{Conclusion}

The observation that neutrino masses are tiny has a natural explanation
in the seesaw model. However, its complicated structure also means that the
neutrino mixing angles do not derive simply from the seesaw components, viz.,
the Dirac matrix $m_D$ and the Majorana matrix $M^{-1}$. In fact, if we
write $m_D=U_0 m_D^{\rm diag} V_0$, $M^{-1}=U_M (M^{-1/2}_{\rm diag})^2 U_M^T$, 
$m_{\rm eff}$ depends, roughly speaking, quadratically on all of the 
components that we displayed.

In this paper, we analyze the problem in several steps.
First, we write $m_{\rm eff}=N N^T$, so that $N$ depends linearly on the 
aforementioned components. But $N$ has the further ambiguity of an arbitrary
RH rotation. We will eliminate this ambiguity by reducing $N$ to the 
triangular form. The lower triangular
form arises naturally if $m_D$ has a hierarchy. However, the upper triangular
form is the easiest to use in order to extract the LH, physical, neutrino
mixing angles. By expressing the triangular matrix elements in terms of RH
rotational invariants, it is easy to transform back and forth between the 
different forms of $N$. Thus, given the parameters in $m_D$ and $M$, we
can deduce the neutrino mixing angles. Conversely, given the physically
plausible values of the neutrino masses and mixing angles, we can obtain
the constraints that must be satisfied by the parameters in $m_D$ and $M$.
Experimentally, we have a fairly good idea about the intrinsic neutrino
parameters. Their masses are most probably hierarchical, the (23) mixing
angle is almost maximal, and the (13) mixing angle is small. From these 
parameters, we can infer the properties of the Majorana masses and the RH
mixing angles of $m_D$, if we assume that $m_D$ is similar to the 
u-quark mass matrix, 
i.e., $m_D$ has small LH mixing angles and $m_i\approx (m_u,m_c,m_t)$. 
It was found that there must be a large hierarchy in $M$,
proportional not only to the ratios of Dirac masses squared, but also to 
the squares of the mixing angles. 
In addition, the RH mixing angles in $m_D$ combined with $M^{-1}$ must 
be very small, and must be equal to the mass ratios in $m_D$. Physically,
the large hierarchy in $M$ implies the existence of intermediate mass
scales. 
It would be most interesting if these conclusions can be corroborated by
other sources.

Using solar neutrino solutions as inputs, we calculated the 
individual Majorana masses. Because of their strong dependence on the 
physical neutrino mixing angles, 
a wide range of values was found. In particular, $M_3$ is so large 
($\gg 4\times 10^{17}~{\rm GeV}$) for the VO solution which makes it 
highly disfavored.
In this work we have not treated the issue of renormalization, although it
can be shown that the RGE effects are small~\cite{RGE}. We have also not 
discussed
the case when either or both of $m_D$ and $M$ are complex. Although
the $2\times 2$ problem can be solved, the $3\times 3$ case does not seem
to have a simple solution. Nevertheless, it can be shown that, 
with hierarchical masses and small angles, complex phases do not 
contribute significantly~\cite{tanimoto}.
We hope to return to this problem in the future.  

\acknowledgements

Our research is supported respectively by DOE grant 
no.~DE-FG02-91ER40681 (T.K.), 
DOE grant no.~DE-FG03-96ER40969 (G.W.), and 
Purdue Research Foundation (S.M.).

\end{document}